\documentclass[lettersize,journal]{IEEEtran}
\usepackage{amsmath,amssymb,amsfonts}
\usepackage{algorithmic}
\usepackage{algorithm}
\usepackage{subcaption}
\usepackage{array}
\usepackage{float}
\usepackage{multirow}
\usepackage{siunitx}
\usepackage{placeins}
\usepackage{textcomp}
\usepackage{booktabs}
\usepackage{soul,color}
\usepackage{stfloats}
\usepackage{url}
\usepackage{hyperref}
\usepackage{verbatim}
\usepackage{graphicx,color}
\usepackage{cite}
\begin{document}

\title{xApp Conflict Mitigation with Scheduler}

\author{IDRIS CINEMRE,~\IEEEmembership{(Member, IEEE)}, TOKTAM MAHMOODI,~\IEEEmembership{(Senior Member, IEEE)}, AMIRMOHAMMAD FARZANEH,~\IEEEmembership{(Member, IEEE)}}
% \affil{Department of Engineering, Faculty of Natural, Mathematical \& Engineering Sciences, King's College, London}
% \affil{Department of Physics, Colorado State University, Fort Collins, 
% CO 80523 USA}
% \corresp{CORRESPONDING AUTHOR: Idris Cinemre (e-mail: idris.1.cinemre@kcl.ac.uk).}
% \authornote{Author Idris Cinmere is funded for his Ph.D. by the Ministry of National
% Education in Turkiye..}
% \thanks{This paper was produced by the IEEE Publication Technology Group. They are in Piscataway, NJ.}% <-this % stops a space
% \thanks{Manuscript received April 19, 2021; revised August 16, 2021.}

% \markboth{xApp Conflict Mitigation with Scheduler}{IDRIS CINEMRE \textit{et al.}}

\markboth{Journal of \LaTeX\ Class Files,~Vol.~14, No.~8, August~2021}%
{Shell \MakeLowercase{\textit{et al.}}: xApp Conflict Mitigation with Scheduler}

% \IEEEpubid{0000--0000/00\$00.00~\copyright~2021 IEEE}

\maketitle

\begin{abstract}

Open RAN (O-RAN) promotes multi-vendor interoperability and enables data-driven control, yet it also gives rise to the challenge of coordinating independently pre-trained xApps whose decisions may conflict. Although the O-RAN architecture mandates offline training and validation to ensure that only rigorously tested models are deployed, dynamic network conditions can still precipitate operational conflicts. These contextual network factors exert a direct influence on conflict occurrence and must therefore be integrated into any resolution mechanism. To this end, a context-aware conflict mitigation framework based on a scheduler approach is proposed in this work, by which competing xApp actions are reconciled through adaptation to real-time network context—without requiring joint training of xApps or further retraining. Through examination of an indirect conflict involving power and resource block allocation xApps and the employment of an Advantage Actor-Critic (A2C) approach to train both the xApps and the scheduler, it is demonstrated that performance is improved by a scheduler trained with contextual variables, relative to independently deployed xApps (i.e., the conflicting case). Moreover, the system attains its highest performance by incorporating baseline xApps and granting the scheduler access to an expanded action space, thereby emphasizing the pivotal role of adaptive scheduling mechanisms. Through these findings, the context-dependent nature of conflicts in automated network management is underscored, since two xApps may conflict under specific conditions yet coexist under alternative scenarios. Consequently, the ability to dynamically update and adapt the scheduler to accommodate diverse operational intents is vital for future network deployments. By enabling dynamic scheduling without necessitating xApp re-training, the proposed framework is positioned to advance practical conflict resolution solutions while ensuring scalability.

\end{abstract}


\section{INTRODUCTION}

The radio access network (RAN)—a critical and costly component of cellular systems—plays a pivotal role in advancing 6G innovation \cite{yungaicela2024misconfiguration}. Once confined to monolithic, hardware-centric architectures, the RAN has evolved into agile and highly customizable frameworks \cite{santos2024managing}, substantially reducing human intervention and bolstering operational efficiency through AI/ML-driven control loops \cite{9913206,10787253}. By supporting zero-touch, multi-vendor ecosystems, these developments lay the groundwork for 6G \cite{10472316, 9397776}, where automation underpins the delivery of diverse, large-scale services. At the same time, heightened autonomy involving multiple independent control loops inevitably leads to scenarios in which individual controllers, each pursuing distinct optimization objectives, may operate in conflict in the absence of effective coordination. To illustrate how such conflicts emerged and were addressed under earlier automation paradigms, Self-Organizing Networks (SON) and Intent-Driven Networking (IDN) are revisited. Building on these historical insights, the potential of Open RAN (O-RAN), a multi-vendor and data-driven architecture, is examined, highlighting its promise for future wireless ecosystems while also magnifying the risk of uncoordinated or conflicting actions among xApps, third-party applications.

One of the earliest large-scale initiatives to integrate autonomous functionalities into cellular networks was the SON, developed under the auspices of the 3GPP \cite{alliance2007ngmn3ggp} and the next generation mobile network (NGMN) \cite{ngmn2007use} alliance to address the growing complexity of 4G systems while simultaneously reducing operational expenses and enhancing service quality \cite{6515045, aliu2012survey, klaine2017survey}. By employing automated closed-loop control processes for self-configuration, self-optimization, and self-healing, SON functions (SONFs) significantly reduce the need for manual intervention in tasks such as neighbor relation management, antenna tuning, and outage recovery \cite{alliance2007ngmn3ggp9}. This approach not only streamlines operational workflows but also sets the stage for more advanced, intelligent automation in the RAN. However, this same autonomy inherently generates potential conflicts whenever multiple SONFs—each targeting the objective of a distinct key performance metric (KPM) (e.g., coverage, capacity, or energy efficiency)—concurrently adjust overlapping network control parameters (NCPs) \cite{alliance20073ggpconflict}.

The conflicts among SONFs can be broadly categorized into four types: NCP, KPM, logical dependency, and measurement conflicts, each rooted in distinct mechanisms of mutual interference \cite{5990691, lateef2013framework, lateef2015lte, bayazeed2021survey}. Frequently observed NCP conflicts arise when multiple SONFs attempt to configure the same NCP (e.g., transmission power) in opposing directions or magnitudes \cite{9201113, moysen2015self}, while KPM conflicts occur when different SONFs influence the same KPM (e.g., throughput) via separate NCPs (e.g., adjusting antenna tilt or handover offsets) \cite{6328322, 7343527}. Additionally, logical dependency conflicts emerge when the outcome of one SONF directly shapes another’s operating conditions \cite{7981545}, and measurement conflicts surface when decision-making relies on outdated or asynchronous data, leading to suboptimal parameter updates \cite{bayazeed2021survey}. 

Early conflict-resolution approaches in SON primarily relied on policy-based strategies that apply static rules to govern SONF actions, including assigning priorities \cite{5990492,6692533}, implementing impact-time sequencing \cite{6328321}, specifying predefined parameter ranges \cite{5683861}, or truncating parameter updates at a defined threshold \cite{6881623}. Although straightforward to implement, these methods lack the flexibility to adapt to evolving network conditions. Meanwhile, co-designing unifies multiple SONFs into a single entity \cite{5073530}, leveraging joint optimization \cite{6328322} or decision trees \cite{6666642}, albeit at the cost of increased complexity. More advanced objective-driven approaches align SONF behavior with high-level operator goals by computing feasible parameter sets in real time, although they must contend with the exponential growth of state spaces and interdependencies among SONFs \cite{6838256,6933337}. Meanwhile, closed-loop frameworks exploit configurable factors such as mutual influences and priorities to fine-tune SONF outputs, but the intricate relationships between NCPs and KPMs remain challenging to model precisely \cite{6139962, 6623198}. In response, recent research has moved toward proactive and cognitive coordination through data-driven algorithms \cite{keshavamurthy2016conceptual, bui2017survey}, including support vector machine (SVM)-based predictions \cite{8335833} and multi-agent reinforcement learning (MARL) \cite{6934885,8406173,8292257}, which enhance the system’s ability to detect and mitigate emerging conflicts under dynamic network conditions.

Building on SON developments, IDN has emerged as a transformative paradigm for 5G RAN management \cite{jeong-nmrg-ibn-network-management-automation-05} by transitioning from rigid, low-level configurations to high-level outcome specifications that reduce operational complexity and foster agile, scalable, and efficient orchestration \cite{9925251, MEHMOOD2023109477, intenrmanandorg}. Recent efforts highlight the potential of IDN to simplify network management in 5G and beyond, including a “one-touch” multi-domain slicing platform that minimizes manual intervention \cite{9443201}, an intent-based orchestrator for dynamically relocating applications in cloud-native 5G networks \cite{9974703}, an NLP-augmented solution for slice provisioning and service deployment tailored to private 5G environments \cite{10097683}, and a slicing approach that leverages mixed integer programming to optimize QoS fulfillment in 6G \cite{10741513}. Although IDN is intended to streamline management and autonomously coordinate multiple RAN operations, it can inadvertently engender conflicts when multiple high-level intents—each encapsulating distinct or overlapping performance objectives and imposing divergent requirements on shared network resources—are deployed concurrently. In \cite{mwanje2022intent}, these conflicts are classified into four principal types: (i) direct parameter conflicts, which occur when separate intents prescribe contrasting adjustments to the same NCPs; (ii) dependency conflicts, arising when interdependent control loops generate circular or inconsistent logic; (iii) contextual conflicts, which manifest only under specific operational conditions (e.g., high user demand or changing mobility patterns) and remain dormant otherwise; and (iv) latent conflicts, encompassing unintended negative outcomes that are not immediately apparent but become evident only upon observing performance degradation or user-experience issues.

To address conflicts in IDN, initial game-theoretic strategies focus on optimizing parameter values via the Nash social welfare function \cite{9829768}, where each KPM is managed by a cognitive function, and direct conflicts are reconciled by selecting a globally efficient equilibrium. This approach is extended with a Fisher market model \cite{9552215} to accommodate varying intent priorities, achieving higher performance and reduced computation time, and further refined through diverse bargaining solutions \cite{10071924} to ensure Jain-based fairness. Meanwhile, penalty-based frameworks model the cost of failing or partially fulfilling each intent’s KPM as a fitness value \cite{9844074}; an evaluation agent then calculates and minimizes the aggregated penalty across all active intents, executing the derived adjustments through closed-loop control. Additionally, a gradient-based approach using the multiple gradient descent algorithm (MGDA) simultaneously minimizes multiple intent-specific loss functions by deriving a unified descent direction, thus offering a systematic way to improve performance across all tasks concurrently \cite{cinemre2024gradient}. More advanced MARL techniques \cite{10001426} allocate resources based on intent-defined priorities (e.g., maximum bit rate), thereby allowing lower-priority objectives to be proportionally degraded in favor of more critical services while maintaining an automated, data-driven resolution of competing requirements.

As it is noted, this persistent challenge of conflict for both SON and IDN frameworks is that independent autonomous optimization functionalities or control loops may inherently conflict in the absence of overarching coordination. Multiple agents, each pursuing distinct yet interdependent objectives, can undermine one another’s performance by independently tuning overlapping parameters or competing for limited resources. This concern becomes even more pronounced with the advent of O-RAN architectures, wherein disaggregated network components and third-party applications (e.g., xApps, rApps).
% —often leveraging distinct training data, optimization targets, and vendor-specific implementations. 
The O-RAN architecture, widely regarded as a cornerstone of future 6G networks \cite{10353004}, accelerates the trend toward disaggregation and open interfaces \cite{polese2023empowering}, enabling multi-vendor interoperability and data-driven control loops. Such openness not only promotes a more diverse ecosystem of third-party xApps but also amplifies the risk of conflict due to heterogeneous training data, optimization targets, and vendor-specific deployments. Consequently, managing these potential conflicts in a scalable, dynamic manner becomes a pressing challenge, particularly under the O-RAN specification that calls for offline training and validation but does not fully prevent operational conflicts under dynamic conditions.

In response to these complexities, this work proposes a context-aware conflict mitigation framework via a scheduler for O-RAN that obviates the need to retrain or co-train xApps explicitly for collaborative operation. Focusing on an indirect conflict scenario involving power allocation and resource block group (RBG) allocation xApps, an Advantage Actor-Critic (A2C) approach is employed to independently train both the xApps and the scheduler. Experimental results demonstrate that even a straightforward scheduler, trained on contextual variables and responsible for selecting the appropriate xApp action, substantially reduces conflicts and improves the total transmission rate compared with independently deployed, potentially conflicting xApps. Furthermore, enabling the scheduler to choose from an expanded pool with baseline xApps achieves the highest overall performance.

The remainder of this paper is organized as follows: Section \ref{sec:background} introduces a brief O-RAN structure, the types of conflicts in O-RAN, surveys existing work on conflict management, identifies key open challenges, and summarizes the motivation for this study. Section \ref{sec:system_model} presents the system model, including an overview of reinforcement-learning techniques from REINFORCE to A2C, and details how the power and RBG allocation xApps are modeled using A2C. Section \ref{sec:secheduler} then describes the proposed scheduler-management framework along with its associated RIC components. Subsequently, Section \ref{sec:performance} provides training results for the xApps, outlines the scheduler-design methods, and discusses simulation outcomes. Finally, conclusions are drawn in Section \ref{sec:conc}.

\section{BACKGROUND}
\label{sec:background}
\subsection{Brief O-RAN Architecture}
\begin{figure}[t]
    \centering
\includegraphics[clip, trim=9cm 10cm 16cm 5cm, width=1\linewidth]{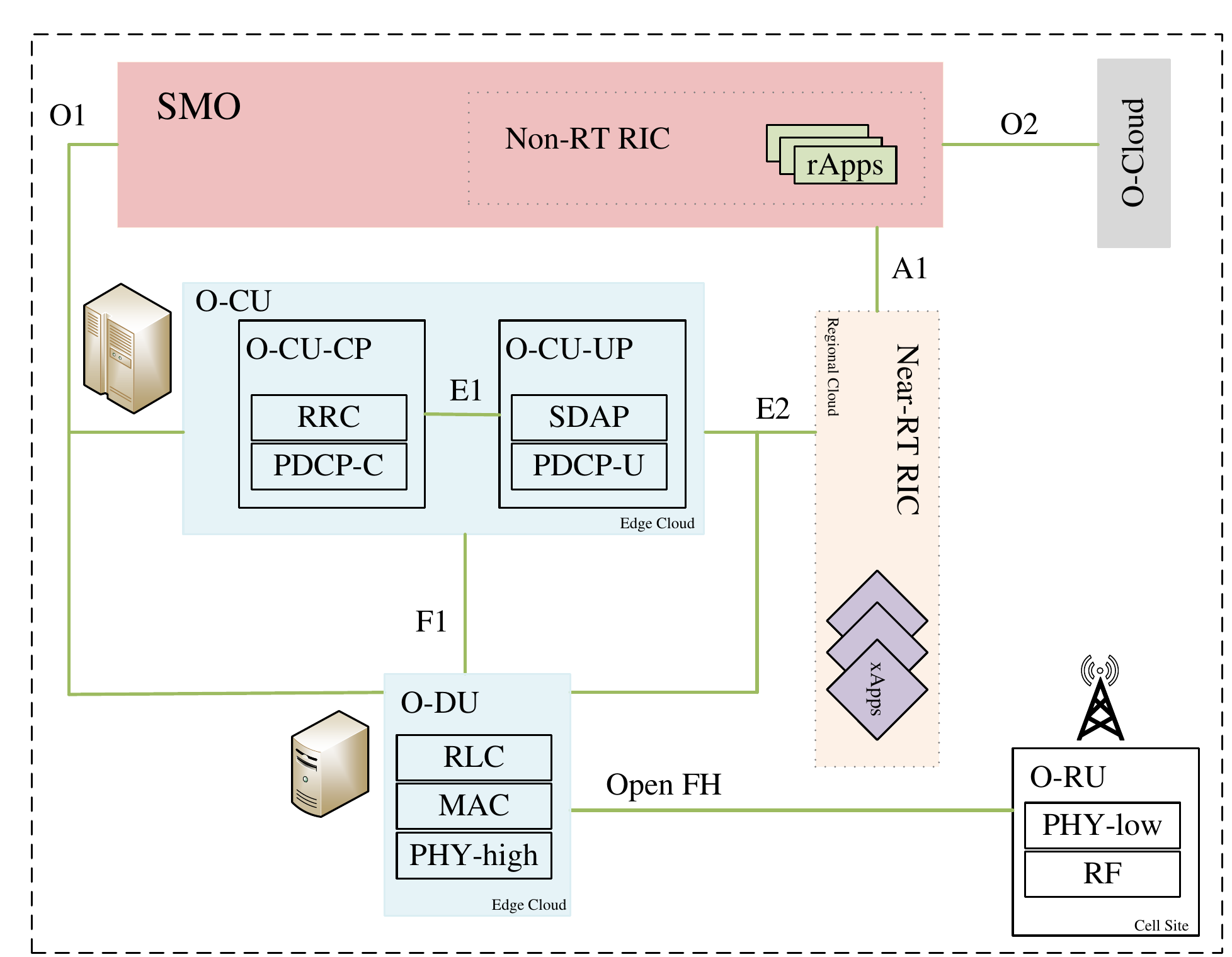}
\caption{An overview of the O-RAN architecture}
	\label{fig:1_brieforan}
\end{figure}

Building upon the 3GPP functional disaggregation framework \cite{NG-RAN}, which partitions the 5G gNB architecture into the remote unit (RU), distributed unit (DU), and centralized unit (CU)—termed O-RU, O-DU, and O-CU in O-RAN nomenclature \cite{thiruvasagam2023open}—the O-RAN architecture delineates these distinct RAN entities (as illustrated in Fig. \ref{fig:1_brieforan}), thereby facilitating multi-vendor interoperability and flexible deployment of the functional components across different hardware infrastructures and network locations \cite{10601697}. Specifically, O-DU, O-CU-CP, and O-CU-UP reside in the edge cloud to satisfy strict latency requirements while the O-RU is physically deployed in the cell site, with the alternative configurations of these functional units to accommodate various scenario requirements, as outlined in \cite{alam2024comprehensive}.

The O-RAN Alliance adopted the $7.2$x functional split to reconcile the simplicity of the O-RU with the data rate and latency constraints at the O-RU/O-DU interface (i.e. open fronthaul (FH)) \cite{10024837} and adopted split option $2$ between the O-DU/O-CU \cite{9839628, yajima2019overview}. The O-CU manages higher-layer non-real-time processes (e.g., packet management) by serving as a logical node to host higher protocol layers, namely RRC, SDAP, and PDCP \cite{motalleb2019joint}. It is functionally divided into the control plane (O-CU-CP), which handles RRC procedures and the PDCP control segment, and the user plane (O-CU-UP), which manages SDAP and the PDCP user segment. These components interact via the E1 interface, thereby enabling flexible scaling and reduced deployment costs for user-plane processing \cite{polese2023empowering, larsen2024evolution}. The O-DU is also a logical node to oversee time-sensitive lower-layer operations and is responsible for hosting the lower-layer protocols of the RAN stack; specifically, the high PHY, MAC, and RLC layers—thereby executing baseband processing tasks such as scrambling, modulation, partial pre-coding, and physical resource block allocation \cite{9695955, ndao2023optimal}. It manages multiple O-RUs over the open fronthaul and provides essential functionalities for data segmentation, scheduling, and multiplexing, with the MAC layer generating transport blocks for the physical layer based on data buffered at the RLC layer \cite{azariah2024survey, 10024837}. This disaggregated design, controlled by the O-CU, enhances deployment flexibility, enables tighter synchronization of lower-layer operations, and allows for more efficient resource utilization. Finally, the O-RU is a physical node dedicated to low-PHY-layer and RF processing; by handling time-domain operations, such as precoding, FFT/IFFT, and cyclic-prefix addition/removal, it remains cost-effective to deploy and maintain at the cell site.

Within O-RAN, as given in Fig. \ref{fig:1_brieforan}, key entities called RAN intelligent controllers (RICs) host intelligent applications that dynamically and efficiently reconfigure the RAN. Specifically, third-party applications, including rApps in the non-real-time RIC (Non-RT RIC) and xApps in the near-real-time RIC (Near-RT RIC), operate independently of the underlying vendor to manage and optimize RAN functionalities under varying latency requirements \cite{10024837}. The Non-RT RIC, operating at time scales longer than $1s$, complements the near-RT RIC’s intelligent RAN operation and optimization by offering guidance, enrichment information (EI), and AI/ML model management. Meanwhile, the Near-RT RIC, deployed at the network edge and operating control loops with periods ranging from $10ms$ to $1s$, hosts xApps—each a microservice specifically designed for radio resource management through specialized interfaces and service models. The Near-RT RIC can be deployed either at the edge or within the regional cloud \cite{10024837, 9579445}, whereas the Non-RT RIC is commonly hosted by the SMO, which manages the configuration and orchestration of RAN components \cite{9695955, thiruvasagam2023open}.
% The Near-RT RIC may be placed in either the edge or regional cloud while the Non-RT RIC is typically hosted in the SMO \cite{10024837, 9695955, thiruvasagam2023open}, which is responsible for the configuration and management of the RAN elements.

The E2 interface connects the Near-RT RIC to RAN nodes for near-real-time monitoring and control loops, while the A1 interface links the Near-RT RIC with the Non-RT RIC to enable policy-based management and facilitate the transfer of EI. Additionally, the O1 interface ensures overarching management and orchestration of O-RAN components, supporting advanced control and automation \cite{ericsson_white_son}. In addition, the O-CU terminates the E2 interface to link with the Near-RT RIC and uses the O1 interface for integration with the SMO framework \cite{alliance2024ran_arch}, enabling advanced control, data-driven optimization, and automation. Meanwhile, the O-DU terminates multiple interfaces, including E2, F1, and open FH, and connects to the SMO via O1 \cite{alliance2024ran_arch}. Moreover, the O2 interface connects the SMO to the O-Cloud, a cloud computing platform that hosts O-RAN functions such as the Near-RT RIC, O-CU-CP, O-CU-UP, and O-DU \cite{bonati2020open}, facilitating the management of both O-Cloud infrastructure and associated workloads \cite{azariah2024survey, 10439167}.

xApps are third-party pluggable modules that extend RAN capabilities \cite{santos2024managing} by implementing specialized, advanced algorithms for radio resource management \cite{9491579, 10278921, zhang2022federated}, slicing \cite{bonati2021intelligence}, interference mitigation \cite{10773839, eskandari2022smart}, and mobility optimization \cite{8690554, orhan2021connection, lacava2023programmable}. Leveraging the standardized E2 interface, xApps can subscribe to performance metrics or event indications and issue corresponding policy-based control directives to the RAN. The Near-RT RIC mediates these interactions, enabling a closed-loop control mechanism in which xApps continuously analyze incoming data and dynamically optimize network behavior (e.g., by adjusting resource allocations or triggering handovers). According to O-RAN specifications \cite{alliance2021ran}, untrained xApps are not permitted; each model must undergo offline training and validation before publication in an AI/ML catalog, thereby mitigating outages or inefficiencies caused by underperforming models. Nonetheless, the concurrent operation of multiple trained xApps can introduce conflicts in network operations, as detailed in the subsequent section.

\subsection{Conflict Types in O-RAN}

\begin{figure*}[t!]
    \centering
\includegraphics[clip, trim=3cm 10cm 17cm 1.5cm, width=1\linewidth]{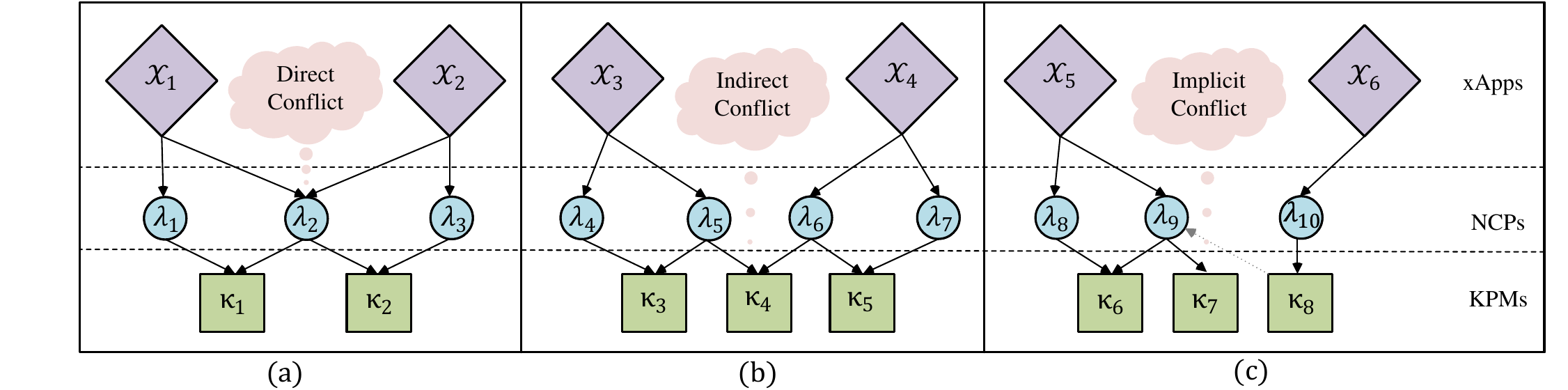}
\caption{Conflict types defined by O-RAN Alliance (WG3)}
	\label{fig:2_conflicttypes}
\end{figure*}

The potential conflict types among xApps were introduced under the Near-RT RIC architecture~\cite{alliance2023conflict}. Recently, the O-RAN Alliance’s working group~3 (WG3) published an independent technical specification on conflict mitigation~\cite{alliance2023conflict2}, analyzing these conflict types and detailing the methods for conflict detection, resolution, and avoidance. This functionality is critical, as multiple xApps, each pursuing distinct objectives, can produce conflicting actions when deployed concurrently.
According to O-RAN technical specifications in ~\cite{alliance2023conflict2}, these conflicts among xApps  (i.e. $\mathcal{X}_1, \mathcal{X}_2, \ldots$) are categorized as follows:

\begin{itemize}
    \item \textit{Direct conflict:} Occurs when multiple xApps request incompatible adjustments to the same configuration of one or more shared NCPs (i.e. ${\lambda_1, \lambda_2,\ldots}$) within the O-RAN. As illustrated in Fig.~\ref{fig:2_conflicttypes}(a),  if two xApps, $\mathcal{X}_1$ and $\mathcal{X}_2$, attempt to modify the same NCP, $\lambda_2$, in opposite directions or with inconsistent magnitudes, a conflict ensues. For example, $\mathcal{X}_1$ may increase transmission power to reduce cell-edge outage, while $\mathcal{X}_2$ may decrease it to mitigate inter-cell interference and enhance capacity. Similarly, $\mathcal{X}_1$ may raise antenna tilt to expand coverage, whereas $\mathcal{X}_2$ may lower tilt to confine signal propagation and reduce interference. 
    
    \item \textit{Indirect conflict:} 
    % Arises when different xApps target different NCPs with the aim of optimizing the same key performance indicators (KPIs) based on their respective objectives within the O-RAN.
    Arises when multiple xApps modify different NCPs to optimize the same KPMs (i.e. $\mathcal{K}_1, \mathcal{K}_2, \ldots$), thereby introducing unintentional interference with each other’s objectives. As illustrated in Fig.~\ref{fig:2_conflicttypes}(b), $\mathcal{X}_3$ adjusts $(\lambda_4, \lambda_5)$ while $\mathcal{X}_4$ adjusts $(\lambda_6, \lambda_7)$ — yet both aim to optimize the same KPM, $\mathcal{K}_4$, in opposite directions or with inconsistent target values. For instance, $\mathcal{X}_3$ might adjust the cell individual offset to balance cell load, while $\mathcal{X}_4$ modifies the cell’s electrical tilt to enhance capacity—both altering the effective cell boundary and thus creating an indirect conflict on the same KPM\cite{10121578}.
    
    \item \textit{Implicit conflict:} Appears when multiple xApps independently optimize distinct KPMs by tuning different NCPs, yet inadvertently impact each other’s objectives. These conflicts often remain hidden because the interdependence between the xApps is not immediately evident. As illustrated in Fig.~\ref{fig:2_conflicttypes}(c), $\mathcal{X}_5$ tunes $(\lambda_8,\;\lambda_9)$ while $\mathcal{X}_6$ modifies $(\lambda_{10})$, each pursuing distinct KPMs. However, $\mathcal{K}_8$ influences one of $\mathcal{X}_5$'s NCPs, thereby implicitly affecting $\mathcal{K}_7$. For example, an Energy Efficiency (EE) xApp may aim to minimize overall power consumption by deactivating certain RUs, while a Slice Management (SM) xApp simultaneously strives to meet high resource demands for bandwidth-intensive applications. These competing goals of reducing energy usage and maximizing throughput generate an implicit conflict \cite{10225786}.

\end{itemize}

\subsection{Related Works}

Existing research on conflict detection and resolution in O-RAN has explored a range of strategies, generally falling into two main categories: \emph{prioritization} and \emph{pre-training}. In~\cite{10225786}, a conflict mitigation (CM) component is introduced within the Near-RT RIC, comprising conflict detection (CD) and conflict resolution (CR) modules. The system leverages manually configured or dynamically learned parameter groups (PGs) to track parameter interactions affecting the same network area. The CD module flags indirect conflicts by comparing newly received E2 control messages against previously allowed configurations in each PG, and the CR module resolves conflicts by prioritizing one xApp’s configuration while rejecting others. Although conceptually straightforward, this approach exhibits limitations, including a simplistic resolution mechanism, potentially cumbersome PG configuration, and the requirement for KPM estimation prior to xApp selection under all conditions. Its extension in~\cite{10121578} subdivides the CD module into three specialized submodules, each addressing a distinct conflict type, and provides further guidance on detecting direct and implicit conflicts.

The study, named as COMIX, outlined in \cite{giannopoulos2025comix} employs a similar conflict detection procedure by defining parameter groups and capturing KPM interactions. As part of the conflict resolution method, each action proposed by the pre-trained xApps is assessed through a scoring function, such as fairness, maximum throughput, or minimum power, using a network digital twin (NDT). However, requiring the NDT to evaluate both actions, and testing every xApp-generated value of NCPs in this manner, could pose significant latency challenges and call into question the necessity of relying on pre-trained xApps. Additionally, the success of the conflict resolution strategy hinges heavily on the performance of the NDT.

A more comprehensive conflict-management solution is provided by the PACIFISTA framework in~\cite{del2024pacifista}, operating within the SMO environment. By interfacing with RICs and RAN nodes to collect profiling data, PACIFISTA dynamically deploys, monitors, and removes xApps/rApps based on predefined conflict-management policies. Unlike earlier methods, it thoroughly evaluates the severity and potential impact of conflicts through statistical analysis of control parameters and KPMs under varied operational conditions (e.g., UE density, signal quality, traffic load). When the severity indexes computed by Kolmogorov-Smirnov, Integral Area, and Chi-Square tests exceed a specified threshold, PACIFISTA either blocks or removes offending applications, thereby prioritizing those that optimize network performance. The principal challenge lies in constructing detailed statistical profiles for each application across multiple operational scenarios and multi-application interactions.

The QoS-Aware Conflict Mitigation (QACM) method proposed in~\cite{wadud2024qacm} addresses conflicts by minimizing the weighted distances of xApp utilities from their QoS thresholds and the squared sum of QoS satisfaction indicators. Deployed as the conflict mitigation controller (CMC) within the CM framework proposed in~\cite{10225786}, QACM expands on Nash’s Social Welfare Function and Eisenberg-Gale solutions~\cite{wadud2023conflict} by incorporating explicit QoS thresholds into its optimization. This design ensures that most xApps meet their QoS requirements, yielding superior performance relative to baseline schemes in both priority and non-priority scenarios. However, the approach assumes reliable KPM prediction which can be challenging in the dynamic nature of RAN conditions where NCPs and dynamic conditions substantially influence KPM behaviors.

Team learning-based methods provide another avenue for conflict mitigation, as demonstrated in~\cite{9838763}, where a deep Q-network (DQN) trains xApps collectively to foster cooperation and eliminate conflicts. While effective, this approach presupposes that all xApps can be jointly trained, which may be infeasible when they originate from different vendors. Building on this idea, the team multi-agent deep reinforcement learning (TMADRL) framework in~\cite{iturria2022multi} extends DQN-based training to multiple xApps controlling different RAN parameters, showing superior performance over sequential and concurrent baselines of MADRL in terms of energy utilization and throughput. However, the method requires careful selection of xApps with a common goal, joint pre-training with cooperative strategies, and significant overhead for action exchange, a challenge that escalates as the number of xApps grows.

The conflict-mitigation technique termed xApp distillation is proposed in \cite{erdol2024xapp}, consolidating multiple ML-based xApps into a single DQN-based controller. Unlike conventional O-RAN conflict mitigation, which discards actions from some xApps, the proposed method also aggregates the experience of all xApps by distilling their policies into a single DQN-based xApp. This student network thus learns a more comprehensive policy from multiple teacher xApps, each of which was pre-trained to manage different or overlapping RAN parameters. The effectiveness of the method depends on teacher quality, and substantial offline data collection for various conditions. Moreover, deploying a single distilled model may limit operational flexibility and require re-distillation if network conditions evolve.

\subsection{Open Challenges and Motivation}

While few solutions have been proposed for detecting and resolving conflicts among xApps (as discussed above), some critical challenges remain underexplored. These challenges must be addressed to ensure robust and adaptive conflict-management mechanisms in real-world O-RAN deployments:

\begin{itemize} \item \textit{Reliance on accurate KPM estimation:} It is overly optimistic to assume that KPMs can be accurately estimated for every possible combination of xApps and their respective actions across diverse network conditions. Consequently, existing prioritization-based approaches that depend on KPM estimation~\cite{10225786, giannopoulos2025comix, wadud2024qacm} remain heavily reliant on the availability of a highly accurate digital twin of the network.

\item \textit{Uncertainty in concurrent xApp deployments:}
In practical environments, accurately predicting which xApps will be concurrently active is inherently challenging. It is unlikely that xApps trained together offline~\cite{9838763, iturria2022multi, erdol2024xapp} will ultimately operate together in real deployments, creating scenarios that deviate from the assumptions made during pre-training.

\item \textit{Immutability of deployed xApps:}
As stipulated in~\cite{10024837, alliance2021ran}, once xApps have been trained, tested, and verified, they cannot be modified after deployment. This immutability poses significant challenges for conflict mitigation, as operators cannot simply retrain or adapt xApp policies (e.g., exchanging actions among xApps by incorporating these actions into the state space, as in~\cite{9838763, iturria2022multi}) to address conflicts that emerge in dynamic network environments.

\item \textit{Context-dependent nature of conflicts:}
In conflict-management scenarios, it is vital to acknowledge that conflicts are context-dependent~\cite{del2024pacifista}. Specifically, varying operational conditions (e.g., user-equipment density, traffic load patterns) can cause two xApps to conflict under certain circumstances while allowing them to coexist under others. However, many existing solutions fail to consider these contextual nuances, resulting in over-generalized or static conflict-management strategies.
\end{itemize}

These observations reveal the shortcomings of current xApp conflict-management strategies and highlight the necessity for more adaptable and context-aware solutions in practical O-RAN deployments. To this end, this work proposes a scheduler-based framework that activates xApps according to explicit contextual variables and user-defined KPM targets, thereby eliminating the requirement to retrain existing xApps. First, we demonstrate the framework through a simulation scenario of \emph{Power} and \emph{RBG Allocation} xApps—each trained using an A2C algorithm—which display a pronounced, context-dependent indirect conflict. To address this issue, we then introduce an intent-driven conflict-mitigation mechanism that constrains the deployable xApps space: the scheduler dynamically selects and orchestrates xApp activations in alignment with prevailing network operational context conditions and specified performance objectives. Finally, we detail two distinct scheduling strategies that exploit contextual inputs and target KPMs derived from user intent, thus offering a flexible and efficient approach to conflict mitigation within O-RAN environments.

\section{SYSTEM MODEL}
\label{sec:system_model}

This study considers an O-RAN cellular architecture with a set of O-RUs, denoted by $\mathcal{B} = \{1, \dots, B\}$, concurrently serving a user set $\mathcal{U} = \{1, \dots, U\}$, managed through the RICs within the O-RAN framework. In this setting, a minimum of $ n \geq 2 $ distinct xApps, denoted as \(\{\mathcal{X}_1, \mathcal{X}_2, \ldots, \mathcal{X}_n\}\), are deployed within the Near-RT RIC to control KPMs in a downlink OFDM scenario. In this model, it is assumed that each O-RU possesses a set of available resource block groups (RBGs), $\mathcal{R} = \{1, \dots, R\}$, which are bundles of consecutive resource blocks as defined in \cite{3gpp2018nr} and represent the smallest allocable time-frequency resources for user assignments.

At any time step \(t \in \{1,2,\dots,T\}\) within episode \(e\), $ \delta_{t,e}^{b, r, u} $ is a binary variable that indicates whether O-RU $ b \in B $ allocates RBG $ r \in R $ to user $ u \in U $, where $ \delta_{t,e}^{b, r, u} \in \{0, 1\}, \, \forall b, r, u $ and $ \sum_{u \in U} \delta_{t,e}^{b, r, u} = 1, \, \forall b, r $. The instantaneous traffic arrivals for each RBG \(r\) allocated to user \(u\) at time step \(t\) in episode \(e\), denoted by \(\varrho_{t,e}^{b,r,u}\), are assumed to follow a Poisson distribution with mean data arrival rate \(d_e\), which remains constant throughout the episode.

Let \(p_{t,e}^{b,r,u}\) denote the power allocated on RBG \(r\) at O-RU \(b\) for user \(u\) during time slot \(t\) in episode \(e\). In a discrete-power setting, each \(p_{t,e}^{b,r,u} \) takes values from a finite set of feasible power levels, $p_{t,e}^{b,r,u}\in\; \mathcal{P}$ where $ \mathcal{P} \subseteq [\,P_{\min}, P_{\max}\,]$. Following the quantization-based discretization in \cite{10375483,8761431}, the set of feasible power levels, $\mathcal{P}$, is given as
\begin{equation}
\label{eq:powerset}
  \mathcal{P}
  \;=\;
  \{\,0\,\}
  \;\cup\;
  \Bigl\{
    P_{\min}\,\beta^k
    \;:\;
    k=0,1,\dots,K-2
  \Bigr\},
\end{equation}
where
\begin{equation}
  \beta
  \;=\;
  \Bigl(\tfrac{P_{\max}}{P_{\min}}\Bigr)^{\tfrac{1}{K-2}},
\end{equation}
and \(|\mathcal{P}|=K\) is the total number of quantization levels, including the \(0\) level to indicate no transmission on the allocated RBG. 

Additionally, the system incorporates user mobility, wherein each user \( u \) moves at a constant random speed \( v_e^u \in [V_{\text{min}}, V_{\text{max}}] \) throughout episode \( e \), characterized by a mean speed \( \mathbb{E}[v_e^u] \), and each user \( u \) has a specified probability \( \rho \) of altering direction \( \Theta_{t,e}^u \) in each time slot \( t \). At any given time slot \( t \) in episode \( e \), the position of user \( u \), denoted by \( (x_{t,e}^u, y_{t,e}^u) \), is updated as:
\begin{equation}
x_{t+1, e}^u = x_{t,e}^u + v_e^u \cos(\Theta_{t,e}^u), \\
\quad 
y_{t+1, e}^u = y_{t,e}^u + v_e^u \sin(\Theta_{t,e}^u).
\end{equation}
The direction at time $ t+1 $ is given episode $e$ as
\begin{equation}
\Theta_{t+1,e}^u = 
\begin{cases} 
\Theta_{t,e}^u, & \text{with probability } (1 - \rho) \\
\Theta_{t,e}^u + \Delta \Theta, & \text{with probability } \rho
\end{cases}
\end{equation}
where $ \Delta \Theta \in [0, 2\pi]$ denotes the change in direction and initial direction of each user at the start of episode \(e\), $\Theta_{0,e}^u$, is drawn from a uniform distribution over the interval \([0, 2\pi]\).
% (i.e. $\Theta_{0,e}^u \sim \mathcal{U}(0, 2\pi)$. 
% This ensures that all possible initial headings are equally likely.

The logarithmic normalized channel state information (CSI) of user $u$ for allocated RBG $r$ from O-RU $b$, denoted by $ \zeta_{t,e}^{b, r, u} $, is defined as

\begin{equation}
\zeta_{t,e}^{b, r, u} = \left\{ \log_{2} \left( 1 + \frac{\sum_{u \in U} \delta_{t,e}^{b^{\prime}, r, u} h_{t,e}^{b^{\prime}, u}}{\sum_{u \in U} \delta_{t,e}^{b, r, u} h_{t,e}^{b, u}} \right) \, \middle| \, b^{\prime} \in B, b^{\prime} \neq b \right\},
\end{equation}
where $ h_{t,e}^{b, u} $ denotes the channel coefficient between O-RU $ b $ and user $ u $ in the position of $ (x_{t,e}^u, y_{t,e}^u) $ and the numerator accounts for interference from all other O-RUs, \(b' \neq b\). Similarly, the signal-to-interference-plus-noise ratio (SINR) for the link between O-RU \(b\) and user \(u\) on RBG \(r\) at time \(t\) in episode \(e\) is
\begin{equation}
\xi_{t,e}^{b, r, u} = \frac{\delta_{t,e}^{b, r, u} \, h_{t,e}^{b, u} \, p_{t,e}^{b, r, u}}{\sum_{b' \in B, b' \neq b} \sum_{u' \in U} \delta_{t,e}^{b', r, u'} \, h_{t,e}^{b', u} \, p_{t,e}^{b', r, u} + \sigma^2},
\end{equation}
where $ \sigma^2 $ is the noise power. Then, the transmission capacity of RBG $ r $ allocated by O-RU $b$ to user $u$ can be formulated as
\begin{equation}
C_{t,e}^{b, r, u} = W^{r} \log_{2} \left( 1 + \xi_{t,e}^{b, r, u} \right), 
\label{eq:7}
\end{equation}
where $ W^{r} $ denotes the bandwidth of RBG $ r $. The transmission rate \( \Psi_{t,e}^{b, r, u} \) equals to \( C_{t,e}^{b, r, u} \) if the length of the queued data exceeds the transmission capacity within a time slot, in which case any data beyond that capacity is assumed to be discarded. Otherwise, (i.e. \( C_{t,e}^{b, r, u} T_s \geq \sum_{u \in U} \delta_{t,e}^{b, r, u} \varrho_{t,e}^{b, r, u} \)), and the transmission rate is given by \( \Psi_{t,e}^{b, r, u} = \varrho_{t,e}^{b, r, u} / T_s \), where \( T_s \) represents the duration of the time slot $t$.

\subsection{From REINFORCE to Advantage Actor-Critic (A2C)}

The REINFORCE algorithm \cite{williams1992simple} is a foundational policy-based method in reinforcement learning (RL) that optimizes the expected return by adjusting the policy parameters in the direction of the policy gradient. 
Formally, the gradient of the expected return with respect to \(\theta\) is:
\begin{equation}
\nabla_{\theta} J(\theta)
= \mathbb{E}_{\pi}\!\bigl[\nabla_{\theta}\log \pi_\theta(\mathbf{a}_t \mid \mathbf{s}_t)\,G_{t}\bigr],
\label{eq:reinforce_gradient}
\end{equation}
where $\pi_\theta(\mathbf{a}_t \mid \mathbf{s}_t)$ is the probability of taking action \(\mathbf{a}_t\) in state \(\mathbf{s}_t\) and $G_t = \sum_{j=t}^{T-1} \gamma^{j-t} \tau_{j}$ denotes the return with $\gamma$ is the discount factor and reward $\tau$ following time step \(t\). This leads to update the policy parameters $\theta$ via
\begin{equation}
  \label{eq:reinforce_update}
  \theta \leftarrow \theta 
  + \eta \sum_{t=0}^{T-1} G_t \,\nabla_\theta \log \pi_\theta(\mathbf{a}_t \mid \mathbf{s}_t),
\end{equation}
where $\eta$ is the learning rate. While this Monte Carlo approach is conceptually straightforward, it exhibits high variance and necessitates waiting until an episode terminates \cite{murphy2024reinforcement}. A common variance reduction technique is to subtract a state-dependent baseline $b(\mathbf{s}_t)$  \cite{sutton2018reinforcement}, which does not depend on the chosen action $a_t$, leading to the unbiased gradient
\begin{equation}
  \label{eq:actorbaseline_gradient}
  \nabla_\theta J(\theta)
  \;\approx\;
  \sum_{t} \nabla_\theta \log \pi_\theta(\mathbf{a}_t \mid \mathbf{s}_t)\,\bigl(G_t - b(\mathbf{s}_t)\bigr).
\end{equation}
Selecting $b(\mathbf{s}_t) = V^\pi(\mathbf{s}_t)$, the state-value function, gives rise to the \emph{Actor-Critic} framework \cite{degris2012model, konda1999actor}. In particular, $V^\pi(\mathbf{s}_t)$ is approximated by a learnable critic $V_\phi(\mathbf{s}_t)$, and one minimizes the mean squared error $\bigl(G_t - V_\phi(\mathbf{s}_t)\bigr)^2$ to train $\phi$, while the policy gradient (actor update) becomes
\begin{equation}
  \label{eq:actorcritic_gradient}
  \nabla_\theta J(\theta)
  \;\approx\;
  \sum_{t} \nabla_\theta \log \pi_\theta(\mathbf{a}_t \mid \mathbf{s}_t)\,\bigl(G_t - V_\phi(\mathbf{s}_t)\bigr).
\end{equation}
Replacing the full return $G_t$ with the one-step return $G_{t:t+1} = \tau_t + \gamma V_\phi(\mathbf{s}_{t+1})$ further decreases variance by bootstrapping from the critic's estimate. This yields the \textit{advantage},
\begin{equation}
  \label{eq:advantage}
  A_t = \tau_t + \gamma V_\phi(\mathbf{s}_{t+1}) \;-\; V_\phi(\mathbf{s}_t),
\end{equation}
as capturing how much better (or worse) the action $\mathbf{a}_t$ is compared to the critic's baseline $V_\phi(\mathbf{s}_t)$ so that the policy gradient takes the form
\begin{equation}
  \label{eq:A2C_gradient}
  \nabla_\theta J(\theta)
  \;\approx\;
  \sum_{t} \nabla_\theta \log \pi_\theta(\mathbf{a}_t \mid \mathbf{s}_t)\,A_t,
\end{equation}
commonly referred to as \textit{Advantage Actor-Critic (A2C)}. The REINFORCE update in (\ref{eq:reinforce_update}) thus becomes
\begin{equation}
  \label{eq:A2C_update}
  \theta \leftarrow \theta
  + \eta \sum_{t=0}^{T-1} \,\bigl[G_{t:t+1} - V_\phi(\mathbf{s}_t)\bigr]\,
    \nabla_\theta \log \pi_\theta(\mathbf{a}_t \mid \mathbf{s}_t).
\end{equation}

In the A2C approach, a synchronous variant of the Asynchronous Advantage Actor-Critic (A3C) algorithm introduced in \cite{mnih2016asynchronous}, a policy network (the \emph{Actor}), $\pi_{\theta}(\mathbf{a}_{t}\mid\mathbf{s}_{t})$, outputs a probability distribution over actions $\mathbf{a}_{t}$ given state $\mathbf{s}_t$ while a value network (the \emph{Critic}), $V_{\phi}$,  estimates the expected return from state $\mathbf{s}_{t}$; $V_{\phi}(\mathbf{s}_{t}) \approx \mathbb{E}[\, G_{t}\mid\mathbf{s}_{t}\,]$. The \emph{advantage function},
$A(\mathbf{s}_{t}, \mathbf{a}_{t}) =  G_{t:t+1} - V_{\phi}(\mathbf{s}_{t})$, compares the discounted one-step return $G_{t:t+1}$ to the Critic's baseline $V_{\phi}(\mathbf{s}_{t})$. By quantifying how much better (or worse) a particular action is relative to the baseline, this formulation reduces the variance of policy gradient updates while maintaining low bias, thereby enhancing learning stability.
The policy gradient update to $\theta$ aims to maximize the expected advantage
\begin{equation}
    \sum_{t=0}^{T-1} 
    A_{t} \;\log \pi_{\theta}\bigl(\mathbf{a}_{t} \mid \mathbf{s}_{t}\bigr),
    \label{eq:policy-grad}
\end{equation}
while the critic update seeks to minimize

\begin{equation}
\label{eq:combined-loss}
\resizebox{\columnwidth}{!}{%
  $\displaystyle
    \mathcal{L}(\theta,\phi)
    = -\sum_{t=0}^{T-1} A_{t}\,\log\pi_{\theta}\bigl(\mathbf{a}_{t}\mid\mathbf{s}_{t}\bigr)
      + \alpha \sum_{t=0}^{T-1} \bigl(G_{t} - V_{\phi}(\mathbf{s}_{t})\bigr)^{2}\, $
}
\end{equation}
where $\alpha>0$ controls the balance between policy and value losses.

% \section{xApps under Coordination of Scheduler}
\subsection{Power and RBG Allocation xApps with A2C}

The power allocation in multi-cell networks has been extensively investigated in the literature using various RL-based methods, including DQN \cite{8761431, 8792117, ahmed2019deep}, DDQN \cite{yang2022dynamic}, A2C \cite{8641248}, A3C \cite{10375483}, TD3 \cite{9771964}, and DDPG \cite{9745785}. The DQN-based power allocation xApp defined in \cite{9838763} is re-designed with A2C method to allocate the optimal power level to each RBG across all  O-RUs by defining input states at time $t$ within episode $e$ as

\begin{equation}
 \mathbf{s}_{t,e} = \Bigl[
\zeta_{t,e}^{b,r,u},\;
\Psi_{t,e}^{b,r,u},\;
p_{t,e}^{b,r,u},\;
\varrho_{t,e}^{b,r,u}
\Bigr]_{\substack{b\in B, r\in R, u\in U}}
\end{equation}
where $\mathbf{s}_{t,e}$ contains four features for every RBGs of each O-RU assigned to the users. The action vector, representing the power allocations for all RBGs, is given by:
\begin{equation}
\mathbf{a}_{t,e}
\;=\;
\Bigl[
p_{t+1,e}^{b,r,u}
\Bigr]_{\substack{b\in B, r\in R, u\in U}}
\end{equation}
where each element of $\mathbf{a}_{t,e}$ is drawn from a finite set \( \mathcal{P} \) defined in (\ref{eq:powerset}). 

In A2C approach, a policy network (the \emph{Actor}), $\pi_{\theta}$, outputs a probability distribution over possible power levels for each RBG per O-RU, while a value network (the \emph{Critic}), $V_{\phi}$, estimates the expected return for a given state. The joint policy of independent action selection per RBG, $a^r$,  

\begin{equation}
\pi_{\theta}(\mathbf{a}_{t,e}\mid\mathbf{s}_{t,e}) 
= \prod_{(b,r,u)} \pi_{\theta}(a_{t,e}^r=p_{t+1,e}^{b,r,u} \mid \mathbf{s}_{t,e}).
\end{equation}
where the categorical distribution over the $K$ power levels for each RBG $r$ is defined as 
\begin{equation}
    \pi_{\theta}(a^r \mid \mathbf{s}_{t,e})
    \;=\;
    \frac{\exp\bigl(z_{r,a^r}(\mathbf{s}_{t,e})\bigr)}
         {\sum_{\iota=0}^{K-1}\exp\bigl(z_{r,\iota}(\mathbf{s}_{t,e})\bigr)},
    \label{eq:actor-softmax}
\end{equation}
% where $z_{r,k}(\mathbf{s}_{t,e})$ is the logit for the $l$-th power level at the $r$-th RBG. 
where $a^r \in \mathcal{P}$ and $z_{r,\iota}(\mathbf{s}_{t,e})$ is the logit for the $\iota$-th power level at the $r$-th RBG. 

The critic, $V_{\phi}(\mathbf{s}_{t,e})$, is a scalar-valued neural network that approximates $\mathbb{E}[\,G_{t,e} \mid \mathbf{s}_{t,e}\,]$, where the discounted return at time step $t$ is
\begin{equation}
    G_{t,e} 
    = \sum_{j=0}^{T-t-1} \gamma^{j}\,\tau_{t+j, e},
    \quad
    \gamma\in(0,1).
    \label{eq:return}
\end{equation}
Here, $T$ is the final timestep within an episode, $e$, and $\tau_{t+j}$ is the reward at time $t+j$.  

Resource allocation in O-RAN has been examined from the perspective of physical resource blocks using a variety of methods; random forest classifier \cite{qazzaz2024machine},  probabilistic forecasting techniques \cite{10266607}, DRL \cite{fiandrino2023explora}, A2C \cite{mollahasani2021dynamic}, DQN \cite{zhang2022federated}.  In this work, the DQN-based radio resource allocation xApp presented in \cite{9838763} is re-designed as an RBG allocation xApp, wherein the A2C method is employed to assign RBGs to users across O-RUs. This procedure parallels that of the power allocation xApp, wherein an \emph{Actor} network outputs a probability distribution over RBG assignment decisions
\begin{equation}
\pi_{\theta}\bigl(\mathbf{a}_{t,e}\,\vert\,\mathbf{s}_{t,e}\bigr)
\;=\;
\prod_{(b,r,u)}
\pi_{\theta}\!\Bigl(a_{t,e}^r=\delta_{t+1,e}^{b,r,u}\,\Bigm|\,\mathbf{s}_{t,e}\Bigr)
\end{equation}
where the state $\mathbf{s}_{t,e}$ and actions $\mathbf{a}_{t,e}$ are defined as 
% and a \textit{Critic} network estimates the expected return from a given state, $\mathbf{s}_{t,e}$,

\begin{multline*}
\label{eq:state_resource_allocation}
\mathbf{s}_{t,e}
\;=\;
\Bigl[
\delta_{t,e}^{b,r,u}\,\Gamma_{t}^{b,r,u},\;
\delta_{t,e}^{b,r,u}\,R_{t}^{b,r,u},\;
\delta_{t,e}^{b,r,u}\,p_{t}^{b,r,u},\; \\
\delta_{t,e}^{b,r,u}\,L_{t}^{b,r,u}
\Bigr]_{\substack{b\in B,\,r\in R,\,u\in U}}
\end{multline*}

% At each time $t$, the xApp decides which RBGs to allocate to which users across all BSs, producing a binary assignment vector
\begin{equation}
\label{eq:action_resource_allocation}
\mathbf{a}_{t,e}
\;=\;
\Bigl[
\delta_{t+1,e}^{b,r,u}
\Bigr]_{\substack{b\in B,\,r\in R,\,u\in U}},
\quad
\delta_{t+1,e}^{b,r,u}\in\{0,1\}.
\end{equation}
At each time step~\(t\), the xApp determines the allocation of RBGs to users for the next time step \(t+1\). Each element $\delta_{t+1,e}^{b,r,u}$ indicates whether O-RU~$b$ assigns RBG~$r$ to user~$u$ at the next time step $t+1$.  \textit{Critic} network estimates the expected return from a given state, $\mathbf{s}_{t,e}$, $V_{\phi}(\mathbf{s}_{t,e}) \approx \mathbb{E}[\,G_{t,e}\mid\mathbf{s}_{t,e}\,]$.

\

\section{SCHEDULER DRIVEN xApp MANAGEMENT FRAMEWORK}
\label{sec:secheduler}

\begin{figure}[t]
    \centering
\includegraphics[clip, trim=8cm 1.5cm 14cm 1cm, width=1\linewidth]{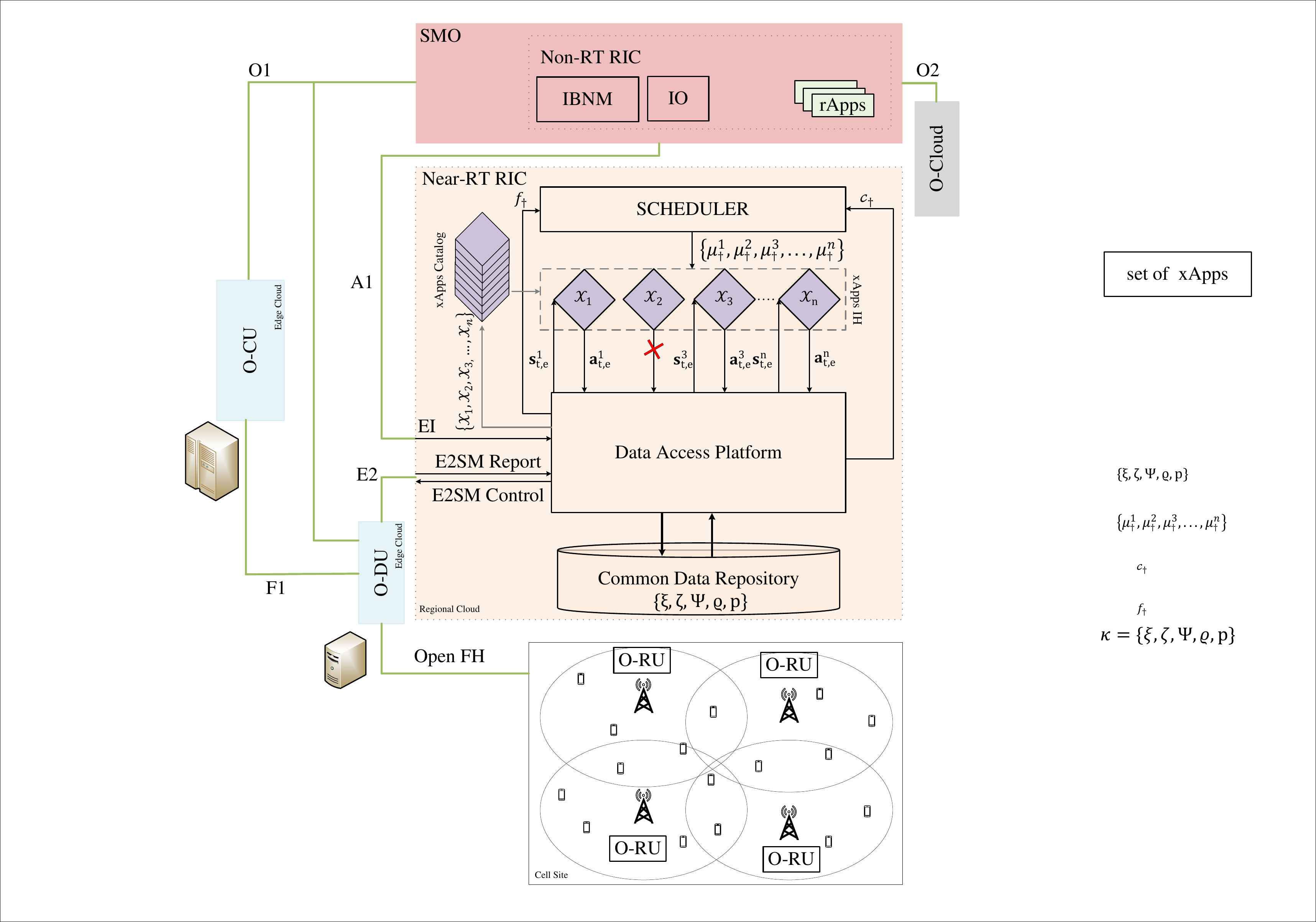}
\caption{Proposed framework with workflow}
	\label{fig:2}
\end{figure}

This study proposes a scheduler within the Near-RT RIC whose primary objective is to coordinate the xApps deployed by the Non-RT RIC in accordance with the intents provided by the MNO, as illustrated in Fig.~\ref{fig:2}.

\subsection{SMO and Non-RT RIC}

The Non-RT RIC facilitates a flexible architecture by offering two primary services \cite{alliance2021ran}: intent-based network management (IBNM) and intelligent orchestration (IO). IBNM enables operators to specify their intents using a high-level language through the Intent Interface \cite{yungaicela2024misconfiguration, 10024837} and converts these high-level intents into the corresponding low-level configuration directives by leveraging large language model (LLM) schemes \cite{10574890, tu2025intent, 10597022, 10588879, 10097683}, aligning with existing SMO practices. The Non-RT RIC parses these intents to deploy the necessary set of rApps and xApps, while IO ensures that xApps and rApps fulfil operator intents and prevent conflicts from multiple applications managing the same KPMs or NCPs \cite{10243548}. Furthermore, leveraging a global network perspective and access to external sources, the Non-RT RIC and SMO relay EI to the Near-RT RIC via the A1 \cite{alliance2021ran_A1}. EI enhances RAN performance by providing data typically unavailable to the RAN, such as capacity forecasts, external data, and aggregate analytics. For example, a rApp within the Non-RT RIC can supply the Near-RT RIC with EI on the predicted evolution of KPMs over the next few seconds for specific users.

\subsection{Near-RT RIC}

According to O-RAN specifications in \cite{alliance2021ran}, the Near-RT RIC incorporates a data access platform that manages data production, sharing, and access among xApps while functioning as middleware between the xApps and a shared data repository, thereby ensuring efficient data storage and distribution. The EI received via the A1 interface is also disseminated to relevant components, including the xApp catalog, to facilitate the deployment of selected xApps to the xApp inference host (IH). The E2 interface enables the Near-RT RIC and its xApps to collect relevant data (i.e., KPMs) from E2 nodes as E2SM reports for the near-real-time control of the RAN. Also, actions taken by xApps to update the NCPs delivered to RAN as E2SM control via E2 \cite{giannopoulos2025comixgeneralizedconflictmanagement}.

\subsection{Scheduler}
 
Upon receiving an intent from the MNO, the Non-RT RIC processes and selects appropriate xApps, \(\{\mathcal{X}_1, \mathcal{X}_2, \ldots, \mathcal{X}_n\}\), according to the target objective of intent, $f_{\dagger}$, (e.g., maximizing data rate) and provides \( \varepsilon \) context variables, \( c_{\dagger}=\left\{c_{\dagger}^{1}, c_{\dagger}^{2}, \dots, c_{\dagger}^{\varepsilon} \right\} \) as EI via A1 for each scheduling period $\dagger$.  The scheduler sends an activation message, \( \left\{ \mu_{\dagger}^1, \mu_{\dagger}^2, \dots, \mu_{\dagger}^n \mid \mu_{\dagger}^n \in \{0, 1\} \right\} \), is to xApps IH, thereby enabling the selected xApps to access the data repository and commence optimization based on their respective objectives.

\section{PERFORMANCE EVALUATION}
\label{sec:performance}
Table~\ref{tab:sim-scenario} presents the parameters of the simulation environment, which was created in Python to facilitate the training and testing of xApps and the scheduler, and which employs an O-RAN mobile network with four O-RUs concurrently serving 16 users.

\subsection{Training of Power and RBG Allocation xApps}

 \begin{figure}[t]
    \centering
    % First subfigure
    \begin{subfigure}[t]{0.53\textwidth}
        \centering
        \includegraphics[clip, trim=1cm 0.1cm 0.1cm 0.1cm, width=1\textwidth]{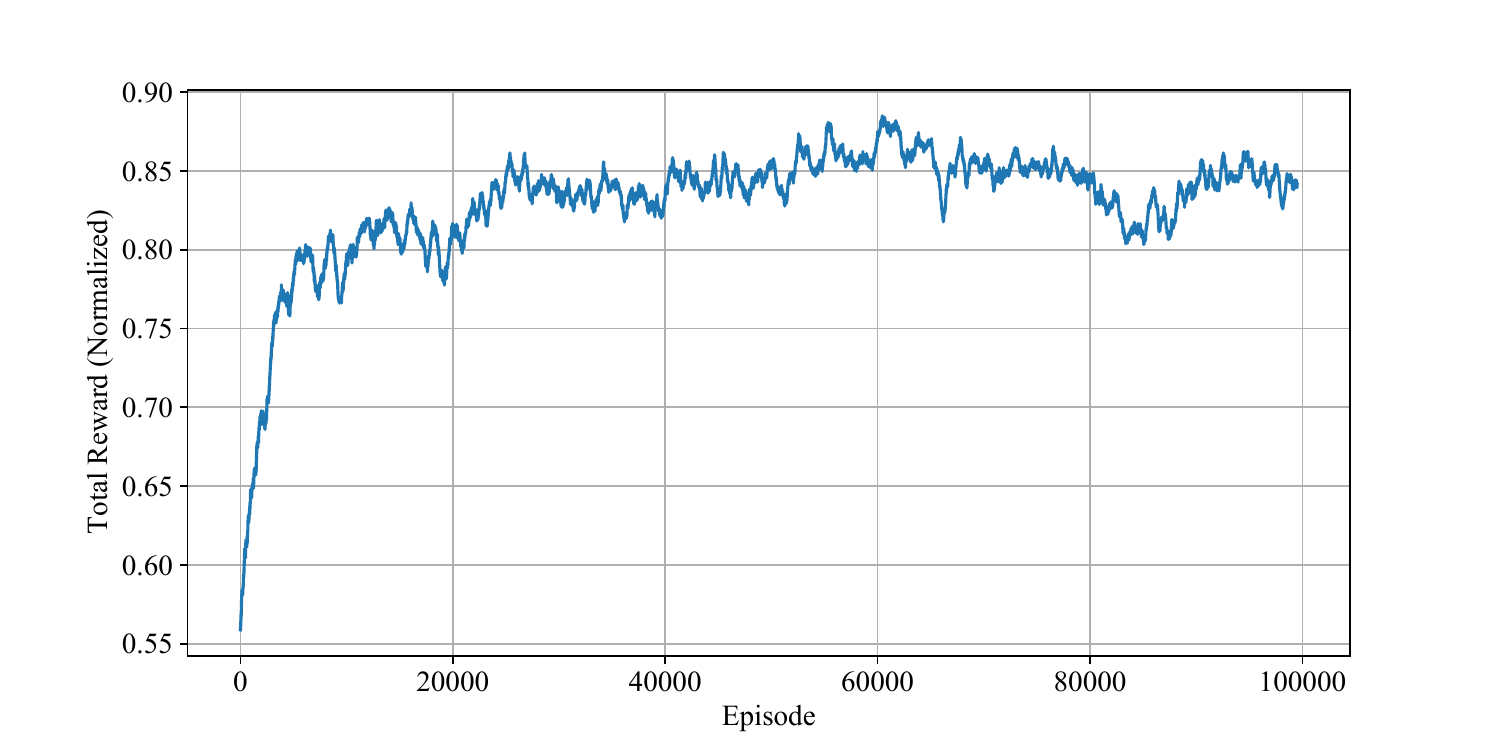}
        \caption{Training performance of Power xApp}
        \label{fig:first}
    \end{subfigure}
    \hfill
    % Second subfigure
    \begin{subfigure}[t]{0.53\textwidth}
        \centering
        \includegraphics[clip, trim=1cm 0.1cm 0.1cm 0.1cm, width=1\textwidth]{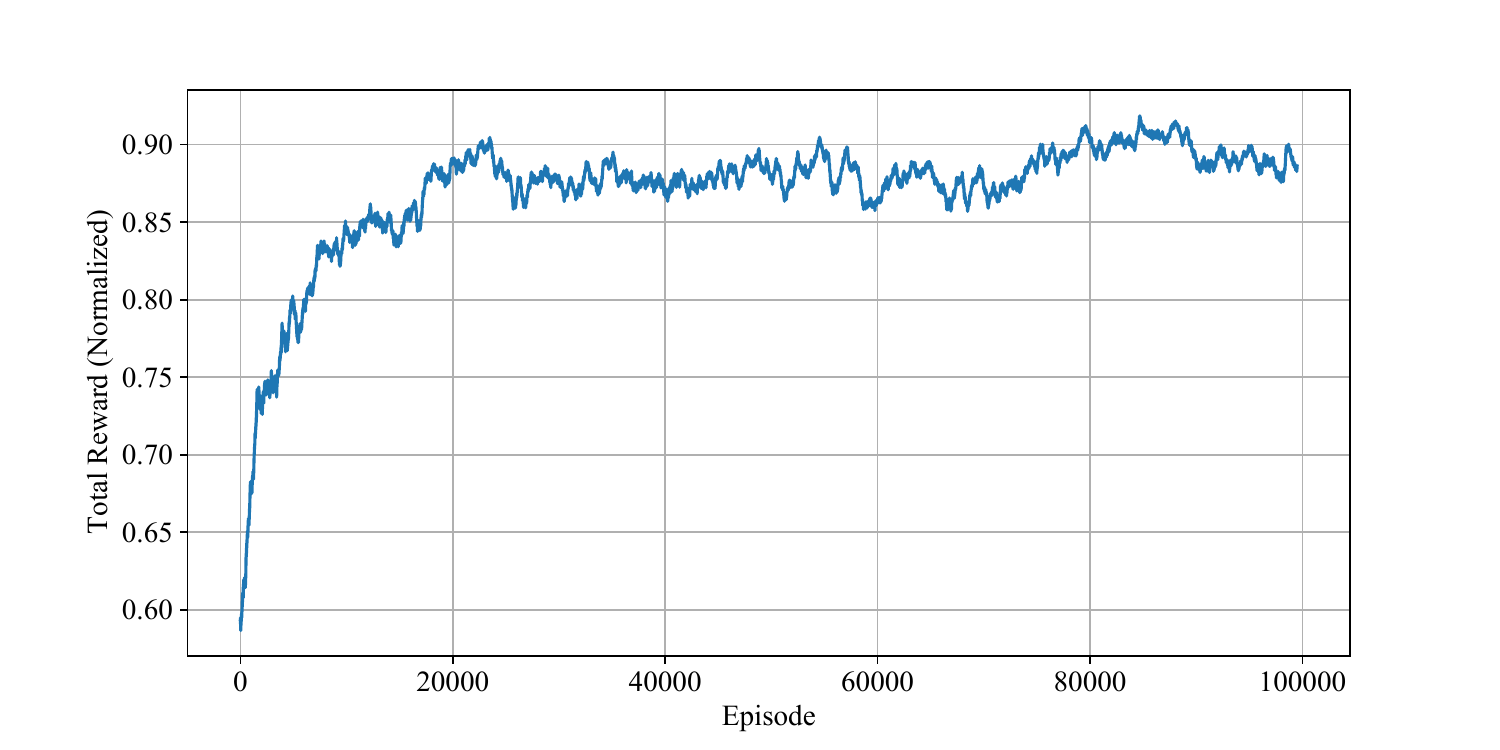}
        \caption{Training performance of RBG xApp}
        \label{fig:second}
    \end{subfigure}
    
    \caption{Moving average (MA) reward of both xApps with a sliding window of 500}
    \label{fig:two_side_by_side}
\end{figure}

 Training of the Power and RBG xApps is conducted in episodes, and the simulation environment is reset at the beginning of each episode (\( t = 0 \)); the users are evenly distributed among four O-RUs, and each user \( u \in U \) is randomly positioned at a distance between $150$ and $450$ m from their assigned O-RU. The propagation environment is modelled by the large-scale fading coefficient of $120.9 + 37.6\log_{10}(\omega) + 10\log_{10}(Z)\;\text{dB},$ where \(\omega\) is the O-RU-to-user distance (in km) and \(Z\) is a log-normal random variable with a standard deviation of \(8\;\text{dB}\). Both xApps are trained under the assumption of equal allocation of all other resources, which serves as a baseline allocation strategy. For instance, the available $12$ RBGs for each O-RU are evenly distributed among connected users during the training of the Power xApp. For each episode $e$, the mean data arrival rate \(d_e\) and the average user speed $ \mathbb{E}[v_e^u]$ are randomly selected from the finite sets $\{3, 5, 7, 9\}$ Mbps and $\{10, 20, 30, 40\}$ m/s, respectively. The objective of both xApps is to maximize the normalized data transmission for each episode, \( \tau_{e} \), calculated as 
\begin{equation}
\label{eq:objectivefunc}
\tau_{e} = \frac{\sum_{t=0}^{T-1} \tau_{t,e}}{d_e \times R \times B}, 
\quad
\tau_{t,e}= \sum_{b=1}^{B} \sum_{r=1}^{R} \sum_{u=1}^{U} \delta_{t,e}^{b,r,u} \Psi_{t,e}^{b,r,u}, 
\end{equation}
where \( \tau_{t,e} \) is the total data rate transmitted across all RBGs ($R \times B$) during each time step $t$ in episode $e$ and the maximum achievable data transmission is defined as the product of the average load \( d_e \) and the total number of RBGs. The training results of the Power and RBG xApps are presented in Fig. \ref{fig:two_side_by_side}, evaluated over $100,000$ episodes, with each episode $e$ comprising $T=50$ time steps.

\begin{figure*}[h]
    \centering
\includegraphics[clip, trim=0.1cm 0.1cm 0.1cm 0.1cm, width=1\linewidth]{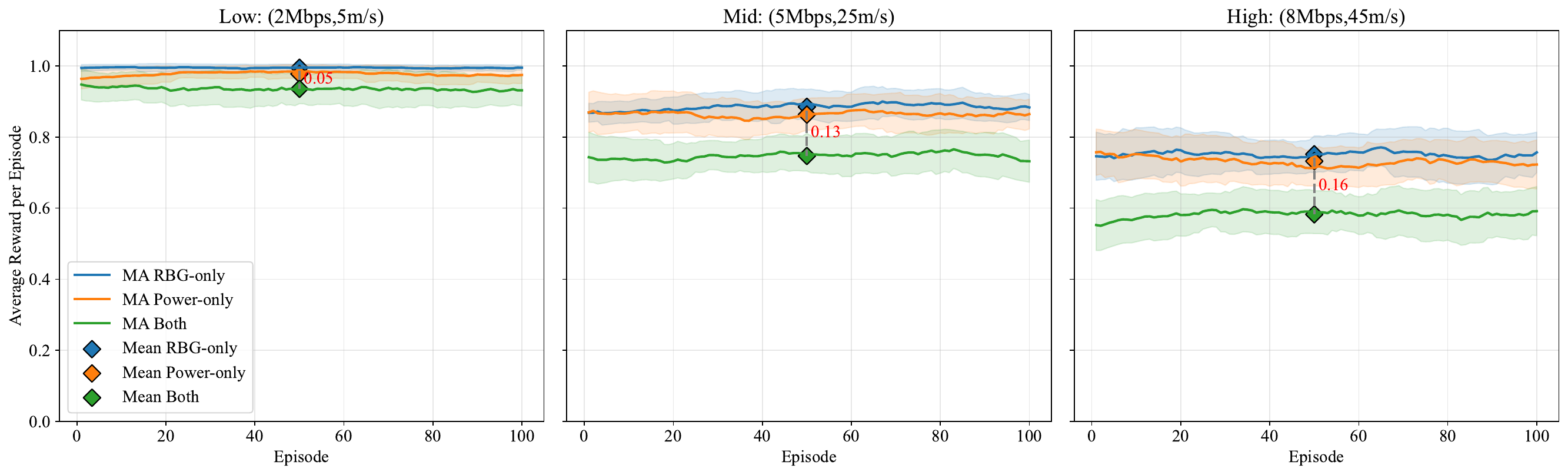}
\caption{Performance comparison of trained xApps and conflicting case}
	\label{fig:5}
\end{figure*}

\begin{table}[t]
\centering
\caption{Simulation Parameters}
\label{tab:sim-scenario}
\begin{tabular}{@{}ll@{}}
\toprule
\textbf{Parameter} & \textbf{Value / Description} \\ 
\midrule
Number of O-RUs ($B$) & $4$ \\ 
Inter-site distance & $900$ m\\
Resource blocks (RBs) & $100$ RBs, each with $12$ subcarriers \\
Resource block groups (RBGs) ($R$) & $12$ RBGs per O-RU \\
Carrier configuration & 20 MHz bandwidth \\
Max transmission power ($P_{min}$) & $38$ dBm \\
Min transmission power ($P_{max}$)& $1$ dBm \\
Additive white Gaussian noise & $-114$ dBm \\
Propagation model & $120.9 + 37.6 \log(\omega)$ dB \\
Log-Normal shadowing & $8$ dB \\
Traffic model of arrival load ($d_e$) &  $\{3, 5, 7, 9\}$ Mbps \\
Number of users ($U$) & $16$ \\
User speed range ($v^u$) & $V_{\text{min}} =1$ m/s, $V_{\text{max}}= 50$ m/s \\ 
Probability of altering direction $(\rho)$ & $0.3$ \\
Training episodes & $10^5$ \\
Time slots per episode ($T$) & $50$ \\
Scheduling period ($\dagger$) & $10$ \\
Slot duration ($T_s$) & $100$ ms \\
Learning rate($\eta$) & $10^{-4}$ \\
Discount factor ($\gamma$) & $0.95$ \\
$\alpha$ in (\ref{eq:combined-loss})  & $0.5$ \\

\bottomrule
\end{tabular}
\end{table}

Fig. \ref{fig:5} illustrates the performance of the trained xApps—specifically, RBG-only, Power-only, and Both (i.e., when deployed independently)—across various scenarios defined by mean data arrival rates (low: $2$ Mbps, mid: $5$ Mbps, high: $8$ Mbps) and average user speeds (low: $5$ m/s, mid: $25$ m/s, high: $45$ m/s). Two primary observations can be drawn from the analysis. First, the independent control of parameters by concurrently deployed xApps, in the absence of coordination, leads to performance degradation across all scenarios. Second, the severity of these conflicts depends on network state variables (i.e., context variables); notably, performance degradation reaches $16\%$ in the high scenario ($8$ Mbps, $45$ m/s), compared to $5\%$ in the low scenario ($2$ Mbps, $5$ m/s). The conflicting behavior of these xApps can be attributed to their independent resource allocation actions. For example, conflicts may occur when the Power xApp assigns a specific power level to an RBG that is not allocated to any user by the RBG xApp. Conversely, conflicts may also arise when the RBG xApp assigns a substantial number of RBGs to a user, while the Power xApp provides a relatively low power allocation for those RBGs.

\subsection{Training of Scheduler}

The scheduler is also modelled using the A2C algorithm, incorporating context variables \( c_{\dagger} \) and reward \( f_\dagger \) as inputs to determine the active xApps for the subsequent scheduling period. In other words, the \emph{Actor} network for scheduler A2C model, $\pi_{\theta}(\mathbf{a}_{\dagger} \mid \mathbf{s}_{\dagger})$, outputs a probability distribution over activation decisions, where the state vector $\mathbf{s}_{\dagger} = \left[ c_{\dagger}^{1},\ c_{\dagger}^{2},\ \dots,\ c_{\dagger}^{\varepsilon},\ f_{\dagger} \right]$ and the action vector $\mathbf{a}_{\dagger} = \left[ \mu_{\dagger}^1,\ \mu_{\dagger}^2,\ \dots,\ \mu_{\dagger}^n \right]$. The initial condition for power and RBG allocation at the beginning of each episode is assumed to be an equal division, consistent with the settings established during the individual training of the xApps. Furthermore, the reward for the scheduler is computed episodically over ${\dagger = 10}$ time steps. Although the scheduler can be configured to optimize multiple objectives, the current study specifically focuses on maximizing the normalized transmission rate, \( \tau_{e} \), achieved by both xApps. Note that only two contextual variables are used; adding more could further improve scheduler performance by capturing a more comprehensive picture of the network’s operational state. This framework eliminates the need for any additional offline re-training or joint training of the xApps themselves; only the lightweight scheduler is re-trained online according to intent and the group of xApps that are employed.

\textit{Method 1: Retain previous action} 

In this method, the training of the scheduler is carried out with two pre-trained A2C xApps— the power (\(\mathcal{X}_1\)) and RBG (\(\mathcal{X}_2\))—to decide which xApp(s) to activate, choosing between individual or simultaneous deployment of these xApps.  The scheduler is designed to make multiple decisions within a single episode (i.e., five decisions per episode (${\dagger} = 10$  time steps)) and share these decisions with xApps through activation messages, $\mu_{\dagger} = \{\mu_{\dagger}^1, \mu_{\dagger}^2\}$ where $\mu_{\dagger}^n \in \{0,1\} \quad \forall\,n=1,2$. Two context variables, the average user speed ($c_{\dagger}^{1}$) and mean data arrival rate ($c_{\dagger}^{2}$), are utilized as the state elements of the scheduler for each scheduling period ${\dagger}$ during the training.
This iterative approach enables the scheduler to refine its activation strategy continuously. Furthermore, if only one xApp is activated at a given decision interval, the system retains the last action made by the previously deactivated xApp, ensuring operational consistency throughout the training. In other words, the scheduler prioritizes the set of xApp actions to be applied to the network and iteratively refines its activation strategy during training. Therefore, the method is regarded as an enhanced, A2C-driven form of prioritization \cite{10225786} and is considered state-of-the-art.
% Consequently, the scheduler prioritizes the set of xApp actions to be applied to the network, refining its activation strategy iteratively during training. Moreover, the method can be regarded as an enhanced form of prioritization \cite{10225786}, with decisions driven by A2C, and is considered a state-of-the-art solution.

\textit{Method 2: Extend with baselines} 

In this method, the scheduler training is conducted with two pre-trained A2C xApps together with two baseline allocation xApps, which distribute resources equally. The scheduler is designed to select one xApp for power allocation (A2C $(\mathcal{X}_1)$ or baseline $(\mathcal{X}_3)$) and one xApp for resource allocation (A2C$(\mathcal{X}_2)$ or baseline $(\mathcal{X}_4)$) to be deployed simultaneously for a given episode, allowing various combinations of baseline and pre-trained models to be activated. This is done by binary activation messages: $\mu_{\dagger} = \{\mu_{\dagger}^1, \mu_{\dagger}^2, \mu_{\dagger}^3, \mu_{\dagger}^4\}$ where $\mu_{\dagger}^n \in \{0,1\} \quad \forall\,n=1,\dots,4$ subject to the constraints \( \mu_{\dagger}^1 + \mu_{\dagger}^3 = 1 \) and \( \mu_{\dagger}^2 + \mu_{\dagger}^4 = 1 \).  In other words, this framework provides the scheduler with the flexibility to dynamically apply either baseline or pre-trained policies based on the current context variables, $ \{c_{\dagger}^{1}, c_{\dagger}^{2}\} $.
\begin{figure*}[t]
    \centering
\includegraphics[clip, trim=0.1cm 0.1cm 0.1cm 0.1cm, width=1\linewidth]{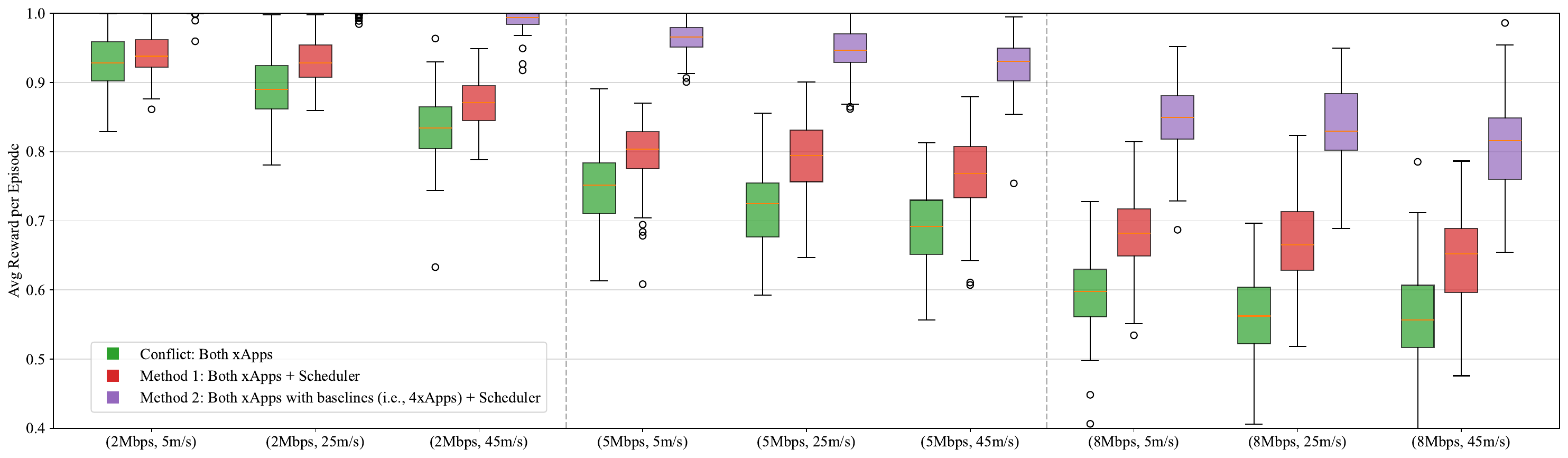}
\caption{Performance comparison of xApps in the presence of a scheduler, based on the reward defined in (\ref{eq:objectivefunc}).}
	\label{fig:6}
\end{figure*}

\textit{Safety Layer: Confidence‑Gated Fallback Mechanism}

Despite extensive training, the A2C scheduler can encounter context vectors that fall outside its training distribution. To bound performance in such regimes, the scheduler can be encapsulated within a lightweight confidence gate: whenever the critic’s value estimate falls below a threshold anomalously, the gate suppresses the RL action and temporarily invokes a pre-determined deterministic fallback policy with a known performance floor. 

Let \(V_{\phi}(\mathbf{s}_{\dagger})\) denote the critic’s value estimate at scheduling instant \(\dagger\) and maintain exponentially weighted estimates of its mean and dispersion with forgetting factor \(\beta\in(0,1]\) \cite{luxenberg2024exponentially}:
\begin{align}
m_{\dagger}   &= (1-\beta)\,m_{\dagger-1} + \beta\,V_{\phi}(\mathbf{s}_{\dagger}),\\
\sigma_{\dagger}&= (1-\beta)\,\sigma_{\dagger-1} + \beta
                   \bigl|V_{\phi}(\mathbf{s}_{\dagger}) - m_{\dagger-1}\bigr|.
\end{align}
A state is classified as \emph{out‑of‑distribution} if the $z$‑score
\[
z_{\dagger} = \frac{V_{\phi}(\mathbf{s}_{\dagger}) - m_{\dagger}}{\sigma_{\dagger}}
\]
falls below a threshold defined by MNO.  When this condition is met, the learned action
\(\mathbf{a}_{\dagger}\) is overridden for the next \(T_{\text{back}}\) decisions by a
deterministic fallback policy \(\pi^{\text{safe}}\):  
(i) equal resource allocation or  
(ii) the single xApp that achieved the highest offline average reward.  
Note that the statistics \((m_{\dagger},\sigma_{\dagger})\) are frozen during the back‑off
window to avoid bias, and the gate re‑arms automatically when the timer elapses.

\subsection{Simulation Results of Testing}

It is assumed that an intent is received as: \textit{"maximize the total transmission rate across all O-RUs"} and the Non-RT RIC selects xApps, \(\{\mathcal{X}_1, \mathcal{X}_2,\mathcal{X}_3, \mathcal{X}_4\}\) based on target, $f$, of the total transmission rate defined in (\ref{eq:objectivefunc}). This set of trained xApps is deployed to xApp IH to be managed by the scheduler. To evaluate the performance of schedulers, tests are explicitly conducted against the conflict scenario, involving the simultaneous deployment of the power and RBG xApps. The evaluations are performed across nine distinct combinations of context variables, specifically \( c_{\dagger}^{1} \in \{2, 5, 8\} \text{ Mbps} \) and \( c_{\dagger}^{2} \in \{5, 25, 45\} \text{ m/s} \), thereby comprehensively capturing diverse system behaviors.
\begin{figure}[b]
    \centering
\includegraphics[clip, trim=0.1cm 0.1cm 0.1cm 0.1cm, width=1\linewidth]{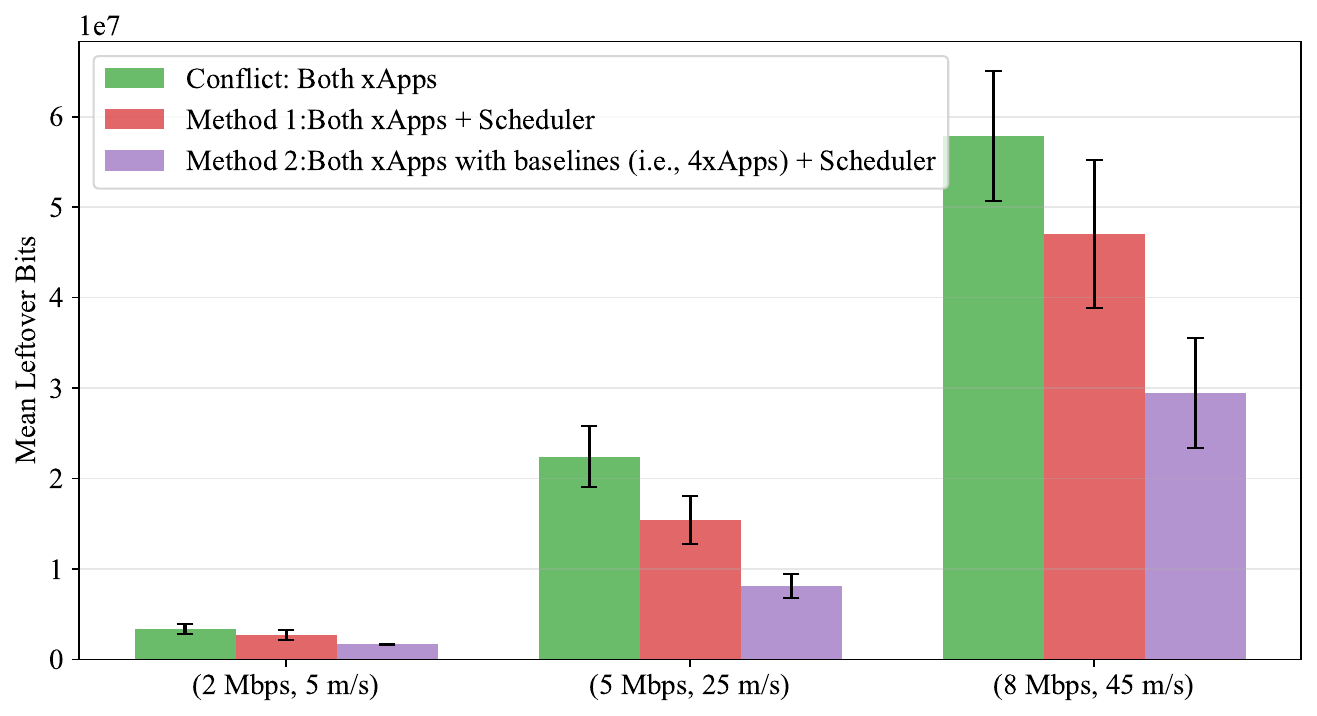}
\caption{Mean leftover bits at the end of each episode}
	\label{fig:7}
\end{figure}

Fig.~\ref{fig:6} compares the performance of multiple active trained xApp variations, including \emph{Both} ($(\mathcal{X}_1)$, $(\mathcal{X}_2)$) deployed independently (i.e. conflicting case); \emph{Method 1: Both} ($(\mathcal{X}_1)$, $(\mathcal{X}_2)$) \emph{with scheduler} and a \textit{Method 2: 4 xApps} ($(\mathcal{X}_1)$, $(\mathcal{X}_2)$, $(\mathcal{X}_3)$, $(\mathcal{X}_4)$) \textit{with scheduler}. It is apparent that variations in the mean user speed have a lesser impact on performance compared to increases in the mean data arrival rate. Furthermore, it is evident that employing the scheduler that retains the previous action (i.e., \textit{Method~1}) results in improved performance relative to the conflicting scenario. However, its effectiveness remains limited due to the constrained action set provided by the independently operating A2C-based xApps. Each xApp determines its actions autonomously, without direct influence from the scheduler, as they are individually optimized to maximize their own objectives by tuning different NCPs. Extending the available action space by incorporating baseline xApps (i.e., \textit{Method~2}) and enabling the scheduler to dynamically choose from this expanded set results in the best overall performance. This framework enhances the scheduler's adaptability by broadening its action space through the integration of baseline actions, enabling flexible selection between pre-trained xApp policies and baseline options based on the prevailing network context.

Furthermore, Fig.~\ref{fig:7} illustrates the mean leftover (i.e., discarded) bits at the end of each episode, arising from insufficient capacity, as defined in (\ref{eq:7}). It is also evident that the degree of the conflict depends on the context variables, and the implementation of scheduler designs contributes to reducing these conflicts and enhancing overall performance. Consistent with Fig.~\ref{fig:6}, the scheduler employing baseline policies in addition to A2C xApps (i.e., \textit{Method 2}) exhibits superior performance compared to the scheduler utilizing exclusively A2C xApps (i.e., \textit{Method 1}).

\section{CONCLUSION}
\label{sec:conc}
In the O-RAN network considered here, multiple distributed O-RUs rely on xApps for essential tasks, such as power and resource block allocations. Despite its capacity for data-driven control, the O-RAN paradigm necessitates robust coordination among pre-trained xApps, as their actions can occasionally conflict.
This work underscores the inherent challenges in estimating KPMs for diverse action combinations of active xApps in the absence of an accurate digital twin, and highlights the context-dependent nature of conflicts in automated network management. Recognizing that pre-trained and validated xApps cannot be modified post-deployment, this study introduces an intent-driven, scheduler-based conflict mitigation framework that selects active xApps based on contextual variables and target KPMs, without necessitating further xApp re-training. Through simulation of an indirect conflict scenario involving Power and RBG allocation xApps, and by leveraging the A2C method for training both the xApps and the scheduler, it is demonstrated that a scheduler utilizing only A2C-trained xApps improves performance relative to independently deployed conflicting xApps. Notably, extending the available set of xApps to include baseline configurations and enabling the scheduler to choose from this expanded pool yielded the most favorable results. These findings advocate for further exploration into adaptive scheduling mechanisms and a broader xApp selection framework to enhance conflict resolution and ensure optimal network performance in practical deployment scenarios through advanced testbed platforms. The ideal scheduler should dynamically manipulate pre-trained xApps based on context-specific network state variables and be easily updated to accommodate various intents and combinations of xApps.

\section*{ACKNOWLEDGMENT}
The authors would like to express their gratitude to Prof. Osvaldo Simeone for his significant contributions to the formulation of the solution presented in this work. Author Idris Cinmere is funded for his Ph.D. by the Ministry of National Education in Turkiye (2018). Also, this work was supported by the Open Fellowships of the EPSRC EP/W024101/1.

\bibliography{ref_res}
\end{document}